\documentclass[a4paper,english,conference]{IEEEtran}
\usepackage{float}
\usepackage{latexsym}

\usepackage{amsfonts}
\usepackage{amsbsy}
\usepackage{amssymb}
\usepackage{times}
\usepackage{graphicx}
\usepackage{setspace}
\usepackage{enumerate}
\usepackage[usenames]{color}
\usepackage[dvips]{pstcol}
\usepackage{epstopdf}
\usepackage[caption=false]{subfig}
\usepackage{cite}
\usepackage{amssymb}
\usepackage{amsfonts}
\usepackage{graphicx}
\usepackage{epsfig}
\usepackage{psfrag}
\usepackage{xcolor}
\usepackage{amsfonts, bm}
\usepackage{epstopdf}
\usepackage{cite}
\usepackage{color}
\usepackage{xcolor}
\usepackage{subfig}
\usepackage{verbatim}
\usepackage{multirow}
\usepackage{booktabs}
\usepackage{amsthm}
\usepackage{makecell}
\usepackage{units}
\usepackage[linesnumbered, ruled]{algorithm2e}
\usepackage{algpseudocode}
\usepackage{amsmath}

\usepackage[linesnumbered, ruled]{algorithm2e}
\usepackage{algpseudocode}
\usepackage{amsmath}

\IEEEoverridecommandlockouts

\usepackage{geometry}
\begin{document}
	
	\title{Digital Wireless Image Transmission via Distribution Matching}
		\author{\IEEEauthorblockN{ Pujing Yang, Guangyi Zhang, and Yunlong Cai}
				\IEEEauthorblockA{College of Information Science and Electronic Engineering, Zhejiang University, Hangzhou, China}
				\IEEEauthorblockA{  E-mail: \{yangpujing, zhangguangyi, ylcai\}@zju.edu.cn}
				\thanks{This work was supported in part by the National Natural Science Foundation of China under Grant U22A2004, and in part by Zhejiang Provincial Key Laboratory of Information Processing, Communication and Networking (IPCAN), Hangzhou 310027, China.  
					
				This paper has been accepted by PIMRC24 workshops, Valencia, Spain, September 2024.
				}  
				}
	
	\maketitle
	\vspace{-3.3em}
	
	\begin{abstract}
		Deep learning-based joint source-channel coding (JSCC) is emerging as a potential technology to meet the demand for effective data transmission, particularly for image transmission. Nevertheless, most existing advancements only consider analog transmission, where the channel symbols are continuous, making them incompatible with practical digital communication systems. In this work, we address this by involving the modulation process and consider mapping the continuous channel symbols into discrete space. Recognizing the non-uniform distribution of the output channel symbols in existing methods, we propose two effective methods to improve the performance. Firstly, we introduce a uniform modulation scheme, where the distance between two constellations is adjustable to match the non-uniform nature of the distribution. In addition, we further design a non-uniform modulation scheme according to the output distribution. To this end, we first generate the constellations by performing feature clustering on an analog image transmission system, then the generated constellations are employed to modulate the continuous channel symbols. For both schemes, we fine-tune the digital system to alleviate the performance loss caused by modulation. Here, the straight-through estimator (STE) is considered to overcome the non-differentiable nature. Our experimental results demonstrate that the proposed schemes significantly outperform existing digital image transmission systems.
		
	\end{abstract}
	
	\begin{IEEEkeywords}
		Digital modulation, joint source-channel coding, wireless image transmission.                                                    
	\end{IEEEkeywords}
	
	\IEEEpeerreviewmaketitle
	
	\section{Introduction}
	Amid the significant increase in data traffic, researchers have dedicated to developing better strategies to meet the demands of multimedia content transmission, such as vision and audio data \cite{Xinchao_Access2024}. Recently, owing to the outstanding information processing abilities of various deep learning models, wireless communication begins to embrace deep learning when designing the systems. In this context, deep learning-based joint source-channel coding (DJSCC) has been rapidly upgraded by combining with deep learning models, especially the autoencoder. DJSCC systems treat the transceiver and the channel as a whole and exploit the deep neural networks (DNNs) to perform encoding and decoding procedures \cite{Huiqiang_TSP2021}. Leveraging the advantages of DNNs, these methods have demonstrated superior performance compared to traditional separation-based systems.
	
	In the realm of wireless image transmission, numerous efforts have been made to integrate deep learning technologies to improve transmission efficiency \cite{Guangyi_TCOM2024}. The main idea is to employ a DNN-based encoder to directly map the input image to channel symbols for transmission, as well as to decode the noisy received symbols with another decoder. In the training phase, the encoder and decoder are jointly optimized with the involvement of channel noise. A pioneer work for this is the DeepJSCC proposed in \cite{Eirina_TCCN2019}, which focuses on leveraging convolutional neural networks (CNNs) to transform the source image into a low-dimensional representation, showing great robustness against noise. Moreover, the advanced Transformer architecture has been considered recently to further enhance transmission performance \cite{Haotian_TWC2024, Yang_ICASSP2023}. To enable adaptive rate adjustment under different channel conditions, the channel signal-to-noise ratio (SNR) has been integrated in the encoding and decoding process, as designed in \cite{ADJSCC,Wenyu_TWC2023,Sixian_JSAC2023}. 
	
	Despite the great performance of these methods, the channel symbols in these existing systems are assumed to be continuous, necessitating analog transmission or full-resolution constellations, which are hard to implement due to the non-ideal characteristics of hardware. As a result, efforts to modulate the channel symbols into discrete constellations are of great significance. To this end, the authors of \cite{JCM} and \cite{Kristy_ICML2019} drew inspiration from the discrete variational autoencoder (VAE), where the system encoder only outputs the probabilities of selecting each constellation. Then, the constellations are generated by sampling from the probability distribution, in which way the output of the encoder is discrete. However, the VAE-based methods are known to lack satisfactory performance in image compression tasks. 
	Other works predominantly concentrate on directly modulating the continuous channel symbols to discrete constellations. Specifically, in \cite{Tze-Yang_ICC2022}, the authors proposed a soft modulation scheme, which serves as a differentiable approximation of the hard quantization operation. Nevertheless, they omit the non-uniform nature of the encoder output, leading to performance degradation.
	\begin{figure*}[!htbp]
		\begin{centering}
			\includegraphics[width=0.82 \textwidth]{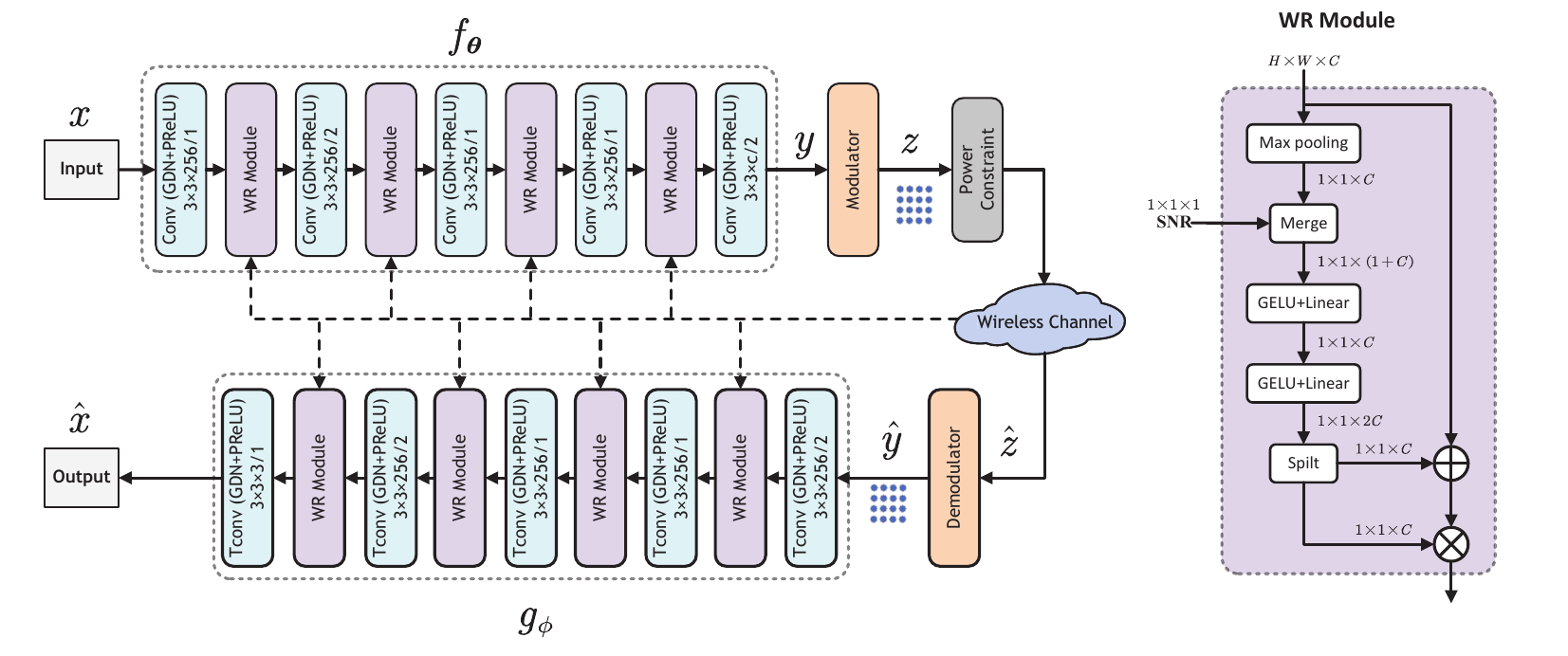}
			\par \end{centering}
		\caption{The architecture of the proposed IDMC. }
		\label{framework}
	\end{figure*}
	
	In this work, we attempt to address these issues to further improve performance for wireless image transmission. To start with, we have identified two main drawbacks of existing methods. First, the output channel symbols in existing methods obey a non-uniform distribution, making it unsuitable to directly modulate with predefined uniform constellations. Second, training the DJSCC model in a discrete manner, e.g., \cite{JCM, Kristy_ICML2019, Tze-Yang_ICC2022}, generally gives rise to a poor representation ability of the encoder \cite{Zongyu_ICML2021}, significantly degrading the overall performance. 	
	To address the first problem, we initially consider a uniform modulation scheme, where the distance between two constellations is adjustable to match the non-uniform nature. More precisely, we update the distance progressively with the optimization of modulation errors.
	In addition, we further design a non-uniform modulation scheme to better match the output distribution. Specifically, we generate the constellations via K-means clustering according to the output distribution of an analog DJSCC system; then the generated constellations are employed to modulate the continuous channel symbols. 
	For both uniform and non-uniform schemes, we fine-tune this digital system to alleviate the performance loss caused by modulation, where the encoder and decoder are initialized with an analog DJSCC. In this way, the second drawback can be overcome. Our simulation results present that the proposed schemes significantly outperform existing digital image transmission systems.

	
	The rest of this paper is structured as follows. Section \ref{SEC2} introduces the framework of our proposed 
	Integrated Digital Modulation and Coding (IDMC). Two modulation schemes for IDMC are presented in Section \ref{SEC3}. Simulation results are presented in Section \ref{SEC4}. Finally, Section \ref{SEC5} concludes this paper.

	\section{System Model}\label{SEC2}
	The proposed IDMC is a digital end-to-end wireless image transmission system, which consists of an encoder, a modulator, a decoder, and a demodulator, transmitting channel input symbols with a finite set of constellations as shown in Fig. \ref{framework}. The input image of size $H\times W\times C$ is represented by a vector $\bm{x} \in \mathbb{R}^{n}$, where $H$, $W$, and $C$ denote the image height, width, and color channels, respectively, and $n=H\times W\times C$. In addition, to adjust the coding strategies, we add an SNR feedback module to enable the encoder and decoder to allocate different resources based on the channel conditions. 
	The encoder $f_{\bm{\theta}}: \mathbb{R}^{n+1}\rightarrow \mathbb{C}^{k}$ maps $\bm{x}$ and channel feedback SNR $\mu$ to a complex vector $\bm{y} \in \mathbb{C}^{k}$, where $\bm{\theta}$ denotes the paremeters of encoder and $k$ denotes the number of transmitted symbols. The encoding process is given by:
	
	\begin{equation}
		\bm{y} = f_{\bm{\theta}}(\bm{x}, \mu) \in \mathbb{C}^{k}.
	\end{equation}
	
	Then $\bm{y}$ is mapped to channel input $\bm{z}$ via modulator $f_{\mathcal{C}}:\mathbb{C}^{k}\rightarrow \mathcal{C}^{k}$, which is represented by:
	\begin{equation}
		\bm{z} = f_{\mathcal{C}}(\bm{y}) \in \mathcal{C}^{k},
	\end{equation}
	where $\mathcal{C} = \{c_1,c_2,...c_M\}$ refers to a finite constellations set and $M$ denotes the modulation order. Due to limited energy in real-world communication systems, we constrain $\bm{z}$ to satisfy an average power constraint before transmission. 
	
	The channel input $\bm{z}$ is then transmitted through a noisy wireless channel, which is denoted by $\eta: \mathcal{C}^{k}\rightarrow\mathbb{C}^{k}$. As the AWGN channel is adopted in our work, this process follows:
	\begin{equation}
		\hat{\bm{z}} = \eta(\bm{z}) =\bm{z} + \bm{n},
	\end{equation}
	where $\bm{n} \sim \mathcal{CN}(0,\sigma^2 \bm{I}_{k \times k})$ is a complex Gaussian vector with variance $\sigma^2$. 
	
	At the receiver, $\hat{\bm{z}}$ is first demodulated to finite constellations using the same set of constellations at the transmitter via the demodulator $g_{\mathcal{C}}:\mathbb{C}^{k}\rightarrow \mathcal{C}^{k}$, which is given by:
	\begin{equation}
		\hat{\bm{y}} = g_{\mathcal{C}}(\hat{\bm{z}}) \in \mathcal{C}^{k},
	\end{equation}	
	where $\mathcal{C}$ is shared by the modulator and demodulator as we will detail in Section \ref{SEC3}, and we will introduce two methods to generate it. Finally the demodulated signal with channel SNR $\mu$ is decoded by the decoder $g_{\phi}:\mathcal{C}^{k}\times \mathbb{R}\rightarrow \mathbb{R}^{n}$ to get the reconstructed image $\hat{\bm{x}}$ as follows:
	\begin{equation}
		\hat{\bm{x}} = g_{\phi}(\hat{\bm{x}}, \mu) \in \mathbb{R}^{n}.
	\end{equation}	
	
	The whole architecture is shown in Fig. \ref{framework}. The encoder and decoder both involve $5$ convolutional layers and $4$ weight reallocation (WR) modules. WR module enables the model to fit more than one channel condition, which effectively avoids storing a large number of models.
	
	The encoder and decoder are jointly optimized to minimize the distortion between the source input $\bm{x}$ and its reconstruction $\hat{\bm{x}}$:
	\begin{equation}
		(\bm{\theta}^*, \bm{\phi}^*) = \mathop{\arg\min}\limits_{\bm{\theta},\bm{\phi}}\mathbb{E}_{p(\bm{x},\hat{\bm{x}})}[d(\bm{x},\hat{\bm{x}})],
	\end{equation}	
	where $d(\bm{x},\hat{\bm{x}})$ is the measured distortion, and $p(\bm{x},\hat{\bm{x}})$ is the joint probability of the source input and its reconstruction. 
	
	\section{Modulation strategy for IDMC}\label{SEC3}
	In this section, we first design two different modulation schemes to match the non-uniform nature of channel input symbols, one generates regular constellations with a learnable constellation minimum distance, while the other generates irregular constellations via clustering. Moreover, we propose a mutual-stage training strategy for IDMC to alleviate the performance loss caused by modulation.
	
	\subsection{Motivation from non-uniform distribution}
\begin{figure}[t]
		\begin{centering}
			\subfloat[]{\label{constellation}\includegraphics[width=4.2cm]{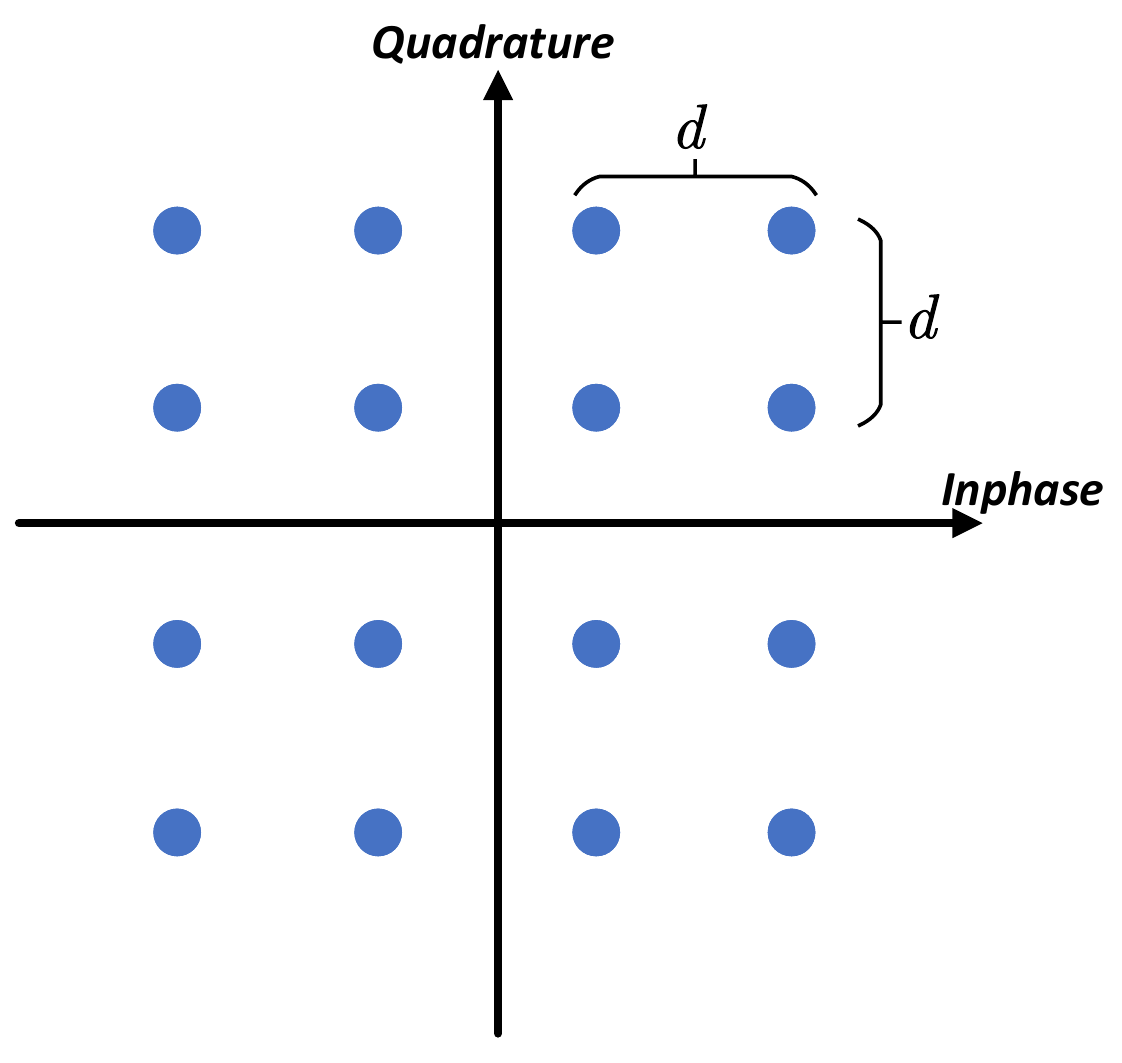}}
			\subfloat[]{\label{distribution}\includegraphics[width=4.2cm]{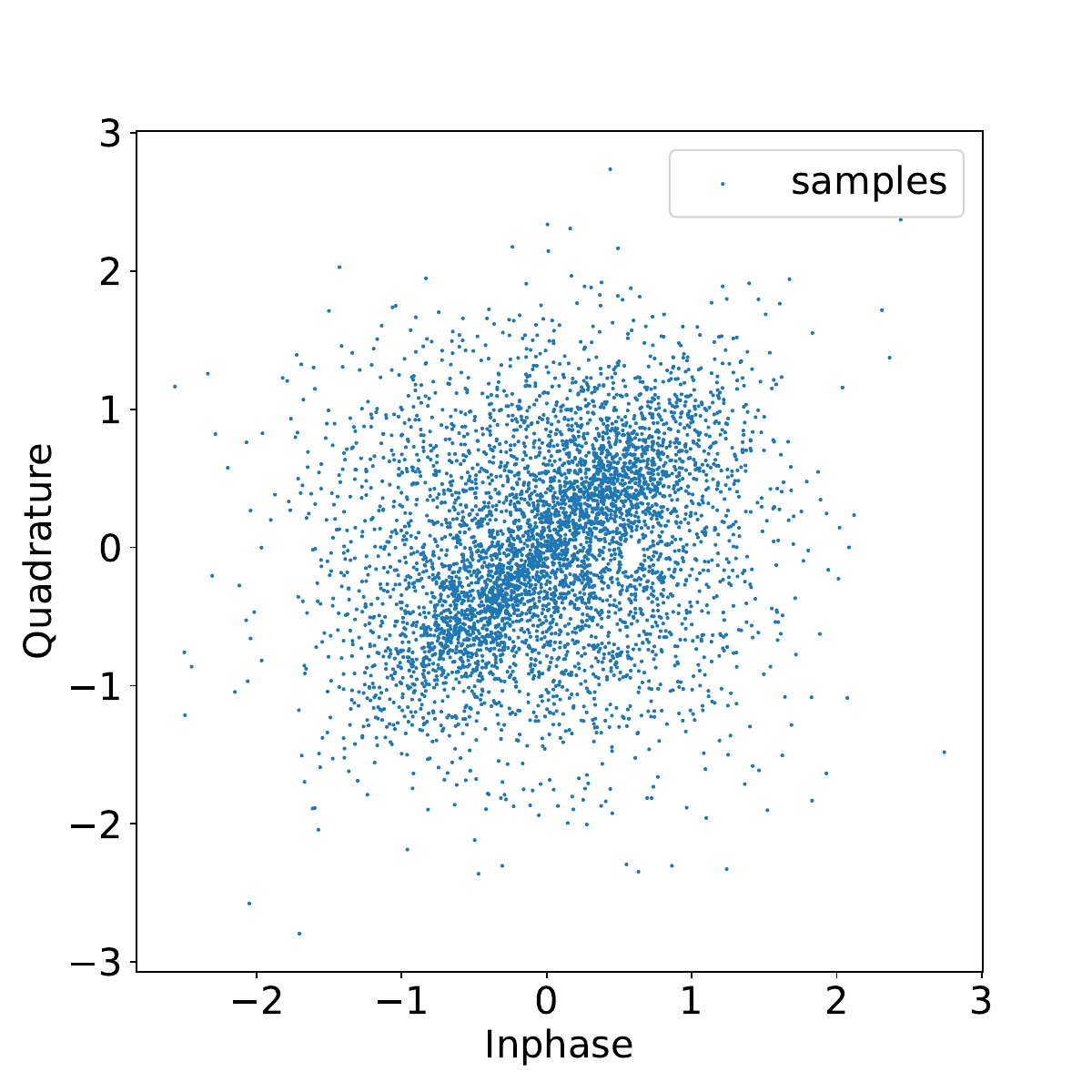}}
			\caption{The uniform constellations and non-uniform channel input symbols. (a) Uniform modulation scheme; (b) The distribution of a sampled image's encoder output symbols.  } 
			\label{cons}
		\end{centering}
	\end{figure}

	To generate discrete constellations from continuous input symbols, we need to design the shape of constellations first. Prior research has predominantly focused on pre-set regular constellations, as illustrated in Fig. \ref{cons}(a). However, the distribution of $\bm{y}$ is not always uniform. To demonstrate this, we randomly sampled an image and plotted its distribution, revealing that some regions are densely populated while others are sparsely occupied, as shown in Fig. \ref{cons}(b). This observation suggests that it is unsuitable to directly modulate with pre-set uniform constellations. Consequently, we propose two different modulation schemes to match the non-uniform nature.
	
	\subsection{Regular constellations generation for IDMC} \label{LSQ}
	Herein, we first consider regular constellations as shown in Fig. \ref{cons}(a). To enhance the model's ability to fit the real non-uniform distribution well, we design a learnable parameter to represent the minimum distance between any two constellations. Inspired by \cite{LSQ}, we refer to the minimum constellation distance as $d$, a trainable parameter in the end-to-end neural network, thus in the forward propagation, we have:
	\begin{equation} \label{equa_round}
		\hat{s}_i = \lfloor clip(s_i / d, B_n, B_p) \rceil \times d,
	\end{equation}
	where $s_i$ is the $i$-th element of $\bm{s}$ and $\hat{s}_i$ is the result of constellation mapping. Here, $clip(s_i / d, B_n, B_p)$ returns $s_i / d$ with values below $B_n$ set to $B_n$ and values above $B_p$ set to $B_p$, and $\lfloor x \rceil$ rounds $x$ to the nearest integer. In this way, we map $\bm{s}$ to integers first and then multiply it by $d$ to eliminate this scaling, realizing regular modulation with a learned constellation distance $d$. For the backward propagation, we copy the gradient of $\bm{s}$ for $\hat{\bm{s}}$ as $\hat{\bm{s}}$ is non-differentiable, specially for $d$, we have:	
	\begin{equation}
		\label{equa_derivative_lsq}
		\frac{\delta \hat{s}_i} {\delta d} =  \frac{\delta s_i} {\delta d} = 
		\begin{cases}
			-\frac{s_i}{d} + \lfloor \frac{s_i}{d} \rceil & \text{if } B_n < \frac{s_i}{d} <B_p, \\
			clip(\frac{s_i}{d}, B_n, B_p) & \text{otherwise,}
		\end{cases}
	\end{equation}

	we note that within the clipping range $[B_n, B_p]$, the derivative is scaled modulation errors, in essence, and the farther away the input symbol is from the constellation, the greater the derivative is. The relationship, aligns with our intuition, indicating that the constellations themselves do not affect the update of the minimum distance $d$ as the modulation errors are $0$ on those constellations. Instead, the farther symbols influence the minimum constellation distance $d$ more as they have larger modulation errors. Hereafter, we abbreviate this IDMC with regular constellations as IDMC-R.

	\subsection{Irregular constellations generation for IDMC} \label{cluster}

	To match the non-uniform distribution of channel input symbols, we initially consider a uniform modulation scheme with a learned constellation distance in Section \ref{LSQ}.
	Furthermore, to achieve a closer alignment with the output distribution, we introduce a non-uniform modulation scheme. 
	
	
	Our strategy employs the K-means clustering algorithm to effectively map continuous symbols into discrete constellations. 
	The constellations are not pre-set but are adapted to the actual distribution of the input symbols, potentially leading to better performance in matching them. Additionally, in contrast to regular modulation schemes that adhere to uniform constellation orders, our method offers a higher degree of flexibility by simply altering the number of clusters, based on the requirements of the system. During constellations generation, we employ the Euclidean distance as the metric for similarity between any two symbols, thus we have:
	\begin{equation} \label{equa_argmin}
		\hat{s}_i =  \mathop{\arg\min}\limits_{c_j \in \mathcal{C}} ||c_j - s_i||^2,
	\end{equation}
	where $||c_j - s_i||^2$ denotes the Euclidean distance between $c_j$ and $s_i$. The entire procedure for irregular constellations generation is outlined in Algorithm \ref{a-kmeans}. In the subsequent text, this IDMC with irregular constellations will be denoted as IDMC-I.
	
	
	\begin{algorithm}[!htbp] 
		\caption{Irregular constellations generation} 
		\label{a-kmeans}
		\SetAlgoLined
		\textbf{Input:} Sampled continuous symbols $\mathcal{S}$, modulation order $M$. \\
		\textbf{Output:} Irregular constellations $\mathcal{C}$. \\
		Randomly initialize $M$ symbols as constellations: $\mathcal{C}=\{c_1,c_2,...,c_k\}$. \\
		Set a flag to record the current state: $flag = 1.$ \\
		\While{flag}{
			$flag = 0$ \\
			\For{$s \in \mathcal{S}$}{
				Assign $s$ to the nearest constellation based on Eq. \ref{equa_argmin}:
				$\hat{s}_i = \mathop{\arg\min}\limits_{c_j \in \mathcal{C}} ||c_j - s_i||^2.$
			}
			\For{$c_i \in \mathcal{C}$}
			{
				Figure out the new constellation $c'_i$: $c'_i = \frac{1}{l} \sum_{j=1}^{l}{s_i}_j$, where ${s_i}_j$ denotes the symbol assigned to $c_i.$
				
			} 
			Reassign $\mathcal{S}$ according to Eq. \ref{equa_argmin}:
			$\hat{s}'_i = \mathop{\arg\min}\limits_{c'_j \in \mathcal{C}'} ||c'_j - s_i||^2.$ \\
			\If{$\hat{s}'_i  \neq \hat{s}_i$}
			{
				$flag$ = 1.
			}
		} 
	\end{algorithm}

	\subsection{Training strategy} \label{train}
	As training the JSCC model in a discrete manner typically degrades the overall performance, we fine-tune the two digital systems from an analog one to alleviate the performance degradation associated with modulation. We develop a mutual-stage training strategy for the two systems:

	\subsubsection{Train an analog model without modulator or demodulator (for both IDMC-R and IDMC-I)}
	For both two modulation schemes, we initiate the training process by developing an analog model that does not incorporate a modulator or demodulator. This serves as our pre-trained model.
	
	\subsubsection{Generate irregular constellations by clustering (only for IDMC-I)}
	Then for IDMC-I, an extra step is implemented compared to the IDMC-R model. This step involves sampling some images from the training dataset. Following this, we extract the encoder output symbols for each sampled image. Subsequently, adhering to Algorithm \ref{a-kmeans}, we execute clustering algorithm on the sampled symbols to derive a finite discrete constellations set. 
	
	\subsubsection{Train the whole model (for both IDMC-R and IDMC-I)}
	\begin{itemize}
		\item For IDMC-R, we add the modulator and demodulator and load the pre-trained model to fun-tune the whole network.
		\item For IDMC-I, once the constellations set is determined, the next step is to integrate modulator, demodulator, and constellations into the model, and then fine-tune the entire network from the pre-trained analog model.
	\end{itemize}
	
	In the forward propagation, no matter which modulation scheme we select, for every symbol to be modulated or demodulated, we choose the nearest constellation as its modulated/demodulated symbol:
	\begin{itemize}
		\item For IDMC-R, the nearest constellation means performing rounding for in-phase and quadrature components, respectively as depicted in Eq. \ref{equa_round}.
		\item For IDMC-I, we need to calculate the Euclidean distance between the continuous symbol and the constellations one by one and select the nearest one, as articulated in Eq. \ref{equa_argmin}.
	\end{itemize}
	
	However, backward propagation presents a challenge at this stage due to the non-differentiability of this operation. Consequently, traditional optimization techniques, including stochastic gradient descent (SGD), prove ineffective, leading to the parameters within the network risk being updated in unforeseen directions. For this, we leverage the straight-through estimator (STE), which substitutes the non-differentiable variables with a differentiable alternative, to facilitate the backward propagation. Thus for backward propagation, we have:
	
	
	\begin{equation}
		\frac{\delta \mathcal{L}}{\delta w} = \frac{\delta \mathcal{L}}{\delta \hat{\bm{s}}} \times \frac{\delta \hat{\bm{s}}} {\delta w} = \frac{\delta \mathcal{L}}{\delta \hat{\bm{s}}} \times \frac{\delta \bm{s}} {\delta w}.
	\end{equation}
	where $\hat{\bm{s}}$ is obtained directly or indirectly from the parameter $w$ within the neural network. Particularly for the learned constellation distance $d$ in IDMC-R, we derive the derivative from Eq. \ref{equa_derivative_lsq}.
	
	As our model is end-to-end jointly learned and optimized, we only need to focus on the difference between $\bm{x}$ and $\hat{\bm{x}}$, thus, we employ the mean square-error (MSE) as our loss function:
	\begin{equation} \label{loss_func}
		\mathcal{L}(\bm{x},\hat{\bm{x}}) = \frac{1}{N}\sum_{i=1}^{N}(\bm{x}_i-\hat{\bm{x}}_i)^2,
	\end{equation}	
	where $N$ denotes the count of sampled images.
	The entire mutual-stage training process is outlined in Algorithm \ref{Training}.

	\begin{algorithm}[!htbp] 
		\caption{Training the IDMC} 
		\label{Training}
		\SetAlgoLined
		\textbf{Input:} Source image dataset $\mathcal{X}$, the learning rate $l_r$. \\
		\textbf{Output:} Parameters $\left(\bm{\theta}^*,\bm{\phi}^* \right)$. \\
		\textbf{First Phase: Train analog model without modulator and demodulator.}\\
		Randomly initialize the parameters $(\bm{\theta},\bm{\phi}).$ \\
			\For{each epoch}{
				Sample $\bm{x}$ from $\mathcal{X}$. \label{step1}\\
				Calculate the loss function based on (\ref{loss_func}). \label{step2}\\
				Update the parameters $\left(\bm{\theta},\bm{\phi} \right)$.\\
			} 
			------------------------------------------------------------ \\
			\textbf{Second Phase (Only for ADCMA-I): Perform clustering on the continuous symbols based on Algorithm \ref{a-kmeans}.}\\
			Update the constellations set $\mathcal{C}.$
			------------------------------------------------------------ \\
			\textbf{Third Phase: Finetune the whole model.}\\
			Load $\left(\bm{\theta},\bm{\phi} \right)$ in the first phase, and particularly for IDMC-I, load $\mathcal{C}$ from the second phase. \\
			\For{each epoch}{
				Repeat step \ref{step1} to step \ref{step2}.\\
				Calculate the gradient through STE. \\
				Update the parameters $\left(\bm{\theta},\bm{\phi} \right)$.}
		\end{algorithm}

		\section{Simulation Results} \label{SEC4}
		In this section, we perform simulations to evaluate the performance of our two proposed models.
		\subsection{Simulation settings}
		\subsubsection{Basic settings}
		We evaluated our IDMC scheme on the CIFAR-10 image dataset, which consists of $60,000$ $32 \times 32 \times 3$ images in total, and $50,000$ of them are training dataset while the rest $10,000$ images are test dataset. All experiments were conducted using Pytorch. The Adam optimization framework was utilized to perform stochastic gradient descent. We initially set the learning rate to $2\times 10^{-4}$ with plans to reduce it after several epochs. We chose $300$ as training epochs and $64$ as batch size.
		
		\subsubsection{Benchmarks}
		We have considered three deep learning-based methods as benchmarks: analog transmission scheme (abbreviated as "Analog"), pre-set regular constellations modulation scheme with STE (referred to "STE"), and another deep learning-based modulation scheme proposed in \cite{JCM} (abbreviated as "JCM"). They are described in more detail below:
		
		\begin{itemize}
			\item "Analog": The network is essentially the same as the one proposed in Fig. \ref{framework}, except that it lacks a modulator and a demodulator. 

			\item "STE": This method also shares a similar structure with that shown in Fig. \ref{framework} except that continuous symbols are modulated using pre-set constellations.
			\item "JCM": It was first introduced in \cite{JCM}. Rather than mapping continuous symbols to discrete symbols directly, it learns the probabilities of mapping to each constellation. 
		\end{itemize}

		\subsection{Clustering Results}
\begin{figure}[t]
			\begin{centering}
				\includegraphics[width=0.42 \textwidth]{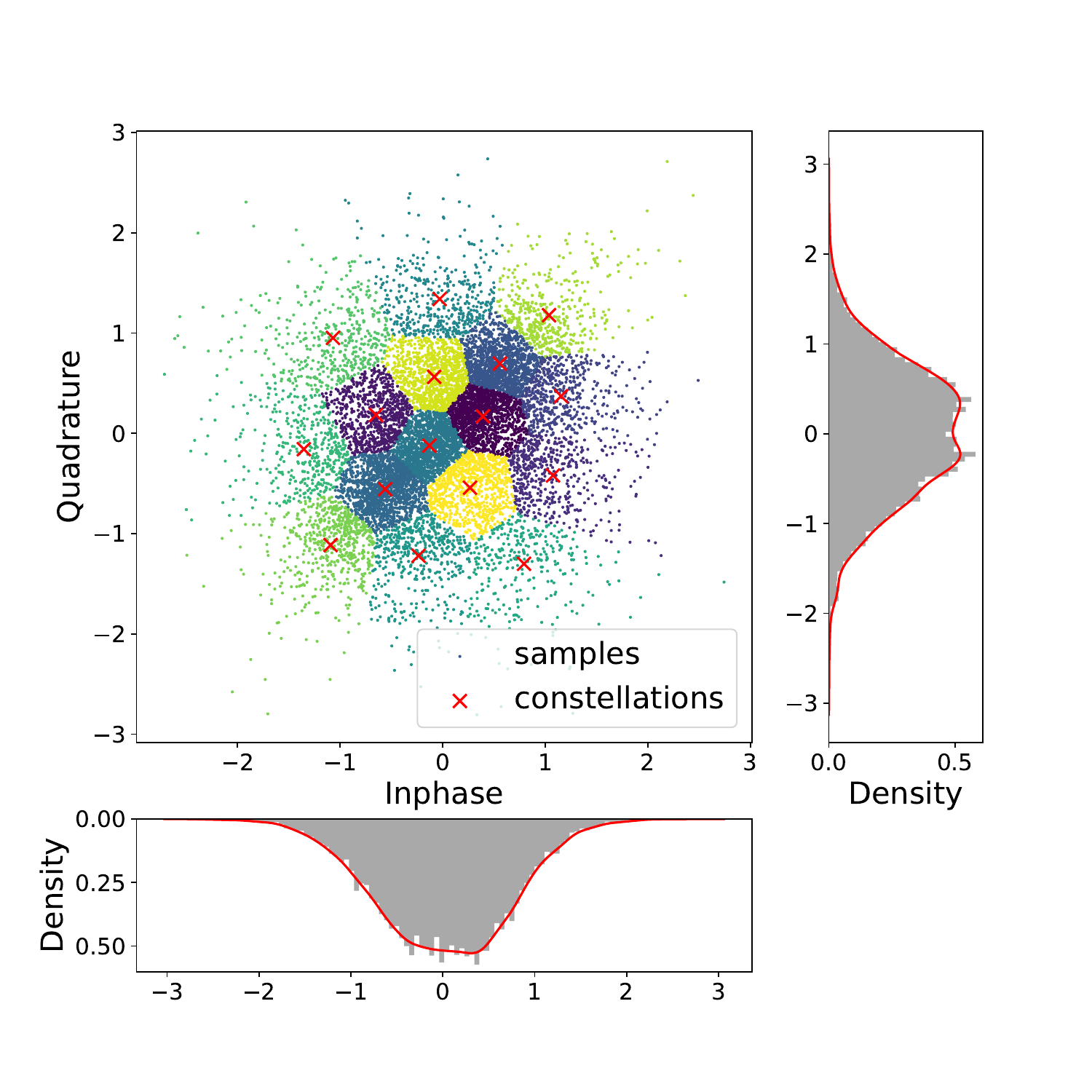}
				\par \end{centering}
			\caption{The distribution of sampled $50$ images' output symbols along with their in-phase and quadrature probability density functions, as well as the clustering result constellations.}
			\label{distribution2}
		\end{figure}

		\addtolength{\topmargin}{0.01in}
		Fig. \ref{distribution2} illustrates the distribution of modulator input along with in-phase and quadrature probability density functions, as well as constellation distribution derived from clustering the continuous input symbols. Our observations reveal that the modulation input distribution is not uniform; indeed, it exhibits a concentration near the center, with a gradual thinning out towards the periphery. Similarly, the clustering results do not display a regular distribution; they are denser at the center and become progressively sparser as we move away from it, closely resembling the distribution of the continuous input symbols. Consequently, this suggests that the constellations can be well-suited to represent these symbols by clustering.
		
		\subsection{Performance comparision}
		The performance of IDMC is quantified in terms of peak signal-to-noise ratio (PSNR). It calculates the ratio of the maximum possible power of the signal to the power of the noise, which is given by:
		\begin{equation}
			\textrm{PSNR} = 10\log_{10}\frac{{\textrm{MAX}}^2}{\text{MSE}}(\textrm{dB}),
		\end{equation}	
		where MAX represents the maximum possible value of an image pixel, and $\textrm{MSE} = \frac{1}{n}||\bm{x}-\hat{\bm{x}}||^2$ is the mean square- error distortion between an original image and its reconstruction. 
		\begin{figure}[t]
			\begin{centering}
				\includegraphics[width=0.42 \textwidth]{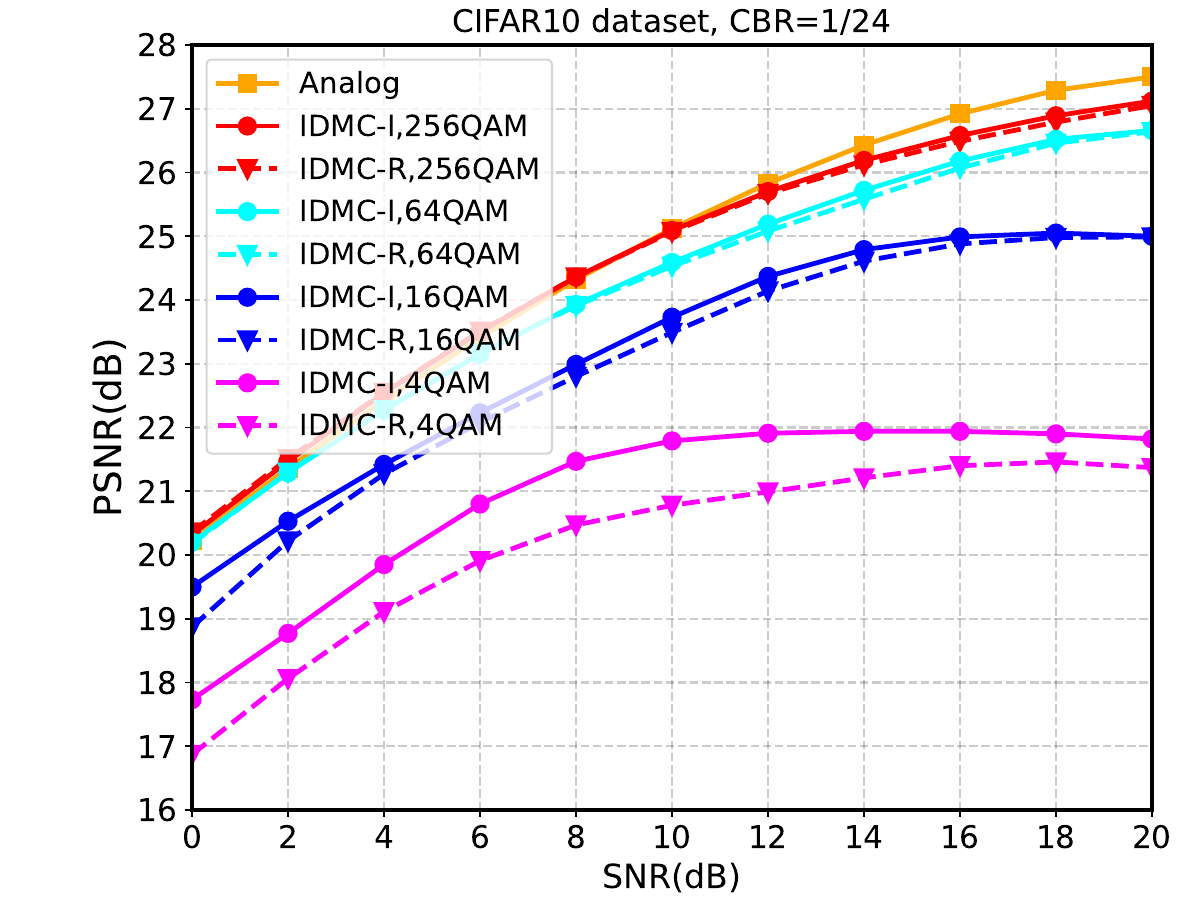}
				\par \end{centering}
			\caption{The performance of the proposed model with different modulation orders versus SNR.}
			\label{cr12_1}
		\end{figure}
		
		\begin{figure}[t]
			\begin{centering}
				\subfloat[]{\label{cr12_2}\includegraphics[width=7.7cm]{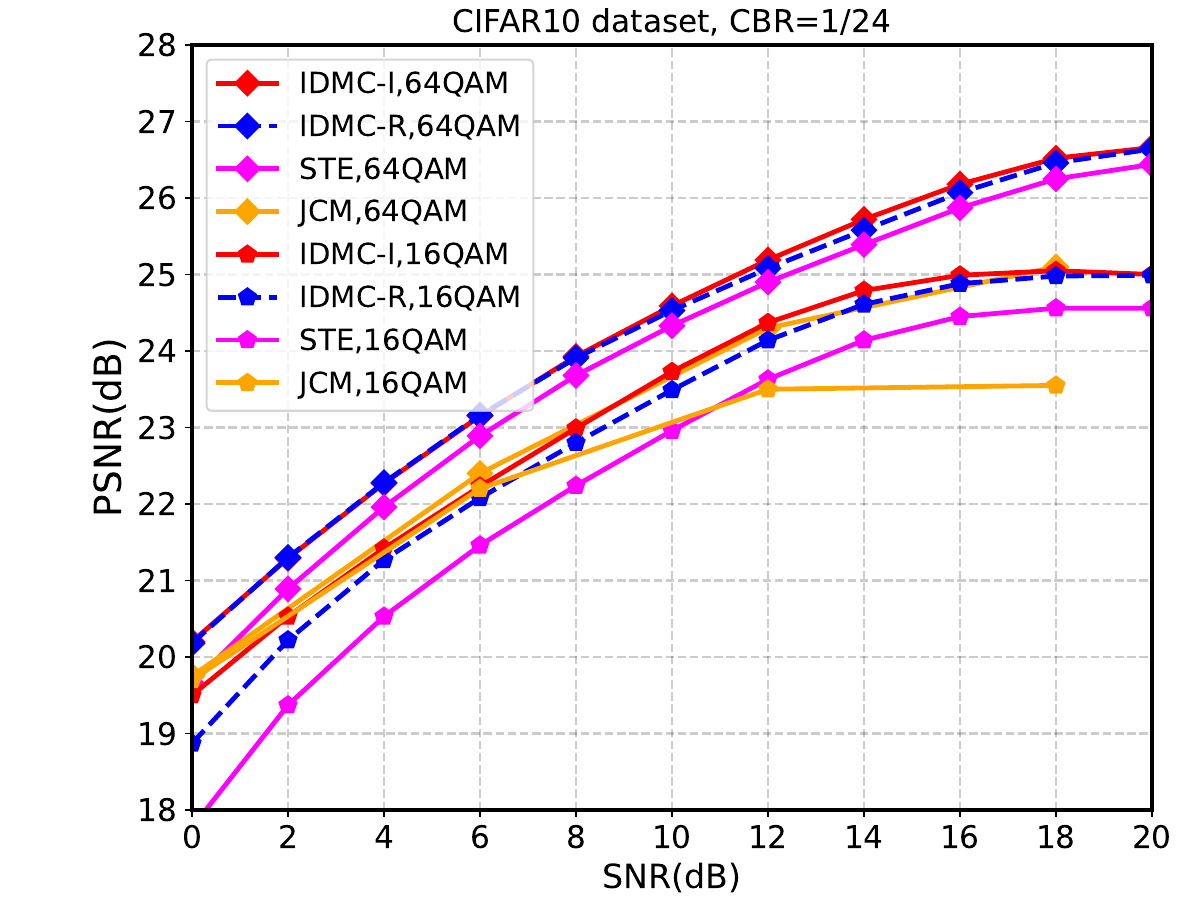}} \\
				\subfloat[]{\label{cr6}\includegraphics[width=7.7cm]{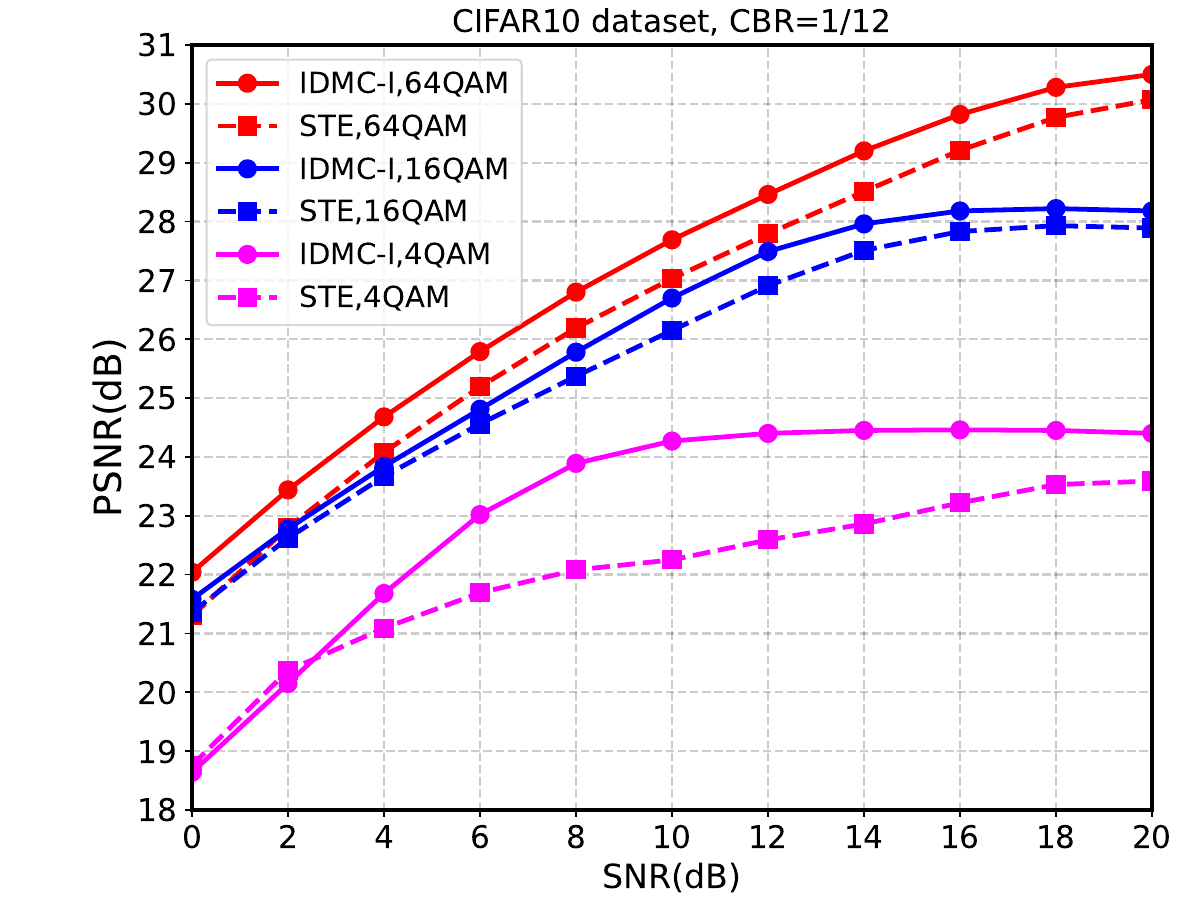}} \\
				\caption{The performance of the proposed model compared with different modulation schemes.} 
				\label{simulation}
			\end{centering}
		\end{figure}	
		We evaluate the performance of our proposed IDMC-R and IDMC-I on the AWGN channel. In all subsequent experiments, the IDMC model is trained on SNR values uniformly distributed from 0 dB to 20 dB and tested within the same range.
		
		Fig. \ref{cr12_1} compares IDMC-R and IDMC-I across various modulation orders at a channel bandwidth ratio (CBR) of $R = 1/24$. Our observations indicate that the performance of both IDMC-R and IDMC-I escalates with an increase in modulation order, approaching "Analog" progressively. This enhancement is attributed to the expanded selection of constellations available at higher modulation orders, which facilitates a reduction in the mean modulation error between a continuous symbol and its discretely modulated representation. Consequently, this mitigates performance loss caused by modulation. 
		We can also observe that the gap between them is particularly larger in low SNRs, where the optimization of the constellation distance in IDMC-R is adversely impacted by noise. Furthermore, IDMC-I exhibits distinct advantages in low-order modulation, showing the great potential of irregular modulation techniques.
		
		
		Figure \ref{simulation} illustrates a comparative analysis of our proposed IDMC-R and IDMC-I schemes against the benchmarks 'STE' and 'JCM'. This comparison spans a range of modulation orders and various CBRs, while also accounting for diverse SNRs. Notably, both IDMC-R and IDMC-I outperform 'STE' and 'JCM' across nearly all SNRs and modulation orders, with a pronounced advantage at 4QAM. These findings underscore the significant potential of our proposed IDMC schemes, with IDMC-I showing particular promise. Furthermore, our analysis reveals that the performance gap between IDMC-I and benchmarks widens at higher CBRs. This is attributed to the increased flexibility in constellation design that higher CBRs afford IDMC-I, thereby enhancing its optimization capabilities.

		\section{Conclusion} \label{SEC5}
		In this paper, we proposed two different modulation schemes for digital image transmission. Considering the irregular distribution of channel input symbols, we first proposed a learned constellation distance scheme with uniform mapping to fit the non-uniformity. Subsequently, to achieve a closer alignment with the irregular distribution, we designed an offline clustering algorithm to generate irregular constellations. Furthermore, we developed a mutual-stage training approach for our model to counteract the performance degradation induced by modulation. Simulation results confirmed the non-uniformity of the distribution of channel input symbols and showed that our proposed modulation schemes significantly enhanced the performance of image transmission. In our future work, we will concentrate on more advanced clustering algorithms and networks to enhance the system's performance. 
		
		\bibliographystyle{IEEEtran}
		\bibliography{IEEEabrv,Reference}
		
	\end{document}